\documentclass[sigconf]{acmart}

\usepackage{booktabs} 
\usepackage{pgfplots}
\usepackage{tikz}
\usepackage{tikz,fullpage}
\usetikzlibrary{arrows,%
                petri,%
                topaths}%
\usepackage{tkz-berge}
\usepackage[position=top]{subfig}
\usepackage{bm}
\usepackage{mathtools}

\setcopyright{rightsretained}

\begin{document}
\title{Learning to Reformulate the Queries on the WEB}

\author{Amir H. Jadidinejad}

\begin{abstract}
Inability of the naive users to formulate appropriate queries is a fundamental problem in web search engines. Therefore, assisting users to issue more effective queries is an important way to improve users' happiness. One effective approach is query reformulation, which generates new effective queries according to the current query issued by users.
Previous researches typically generate words and phrases related to the original query. Since the definition of query reformulation is quite general, it is completely difficult to develop a uniform term-based approach for this problem.
This paper uses readily available data, particularly over one billion anchor phrases in Clueweb09 corpus, in order to learn an end-to-end encoder-decoder model to automatically generate effective queries. Following successful researches in the field of sequence to sequence models, we employ a character-level convolutional neural network with max-pooling at encoder and an attention-based recurrent neural network at decoder. The whole model learned in an unsupervised end-to-end manner.
Experiments on TREC collections show that the reformulated queries automatically generated by the proposed solution can significantly improve the retrieval performance.
\end{abstract}

%
%
%



\maketitle

\section{Introduction}
\label{sec:introduction}

    Different queries are possible for a given information need. Formulating the appropriate query for web search can be a difficult task, given that users may not be aware of the right vocabulary to use.
    Previous researches~\cite{qru12,ir_verbose_queries15} show that query performance can be improved by more than 60\%, if the best query formulation is used for each query~\footnote{QRU-1~\cite{qru12} is a public dataset that contains reformulations of TREC queries as available in query logs of a commercial web search engine.}.

    To tackle this issue, extensive research works have been conducted to help users issue more effective queries. There are two major directions in this research, namely, \textit{query suggestion (recommendation)} and \textit{query reformulation (refinement, modification or rewriting)}~\cite{qr_joint_modeling15}. In query suggestion, previous queries are recommended to the new user according to the closeness with the user's original query. Therefore, query suggestion techniques cannot provide new queries which are not visited in the query log~\cite{when_query_suggest13}. Previous research~\cite{qr_joint_modeling15} show that about 70\% of the satisfactory queries are previously unseen queries in web search.

    On the other hand, in query reformulation, the original query is modified with the aims to solve the vocabulary mismatch problem in information retrieval. The new generated query have a better match with relevant documents. Since the definition of query reformulation is quite general, and includes, among others, abbreviation induction, term expansion, term substitution and term reduction~\cite{ir_verbose_queries15,qr_strategies09}, it is completely difficult to develop a uniform term-based approach for this problem.

    In this paper, we focus on the latter task, namely web query reformulation. Despite considering term-based approaches, we focus on a model which can learn to reformulate the input queries on the web.
    Given a query $q$, we wish to learn a model to generate a modified query $q^{r}$ with the aim of better matching the vocabulary of the relevant documents. We will write this: $q \xmapsto[]{} q^{r}$ and refer to $q^{r}$ as a reformulation for input query $q$.

    In recent years, Deep Neural Networks (DNN) have yielded significant performance improvements on speech recognition, computer vision and natural language processing tasks~\cite{neural_ir16}. The application of these new machine learning approaches in addressing traditional and emerging information retrieval tasks is an emerging hot research topic, so-called \textit{Neural Information Retrieval}~\cite{neural_ir16} or \textit{Neu-IR}~\cite{neu_ir16}.

    Deep learning models require rich sources of training samples. web search engines naturally accumulate a lot of log data, including submitted queries, viewed search results, and clicked URLs.
    In absence of query log, previous researches~\cite{query_reformulation_anchor10} revealed that using anchor text as a simulated query log is as least as effective as a real log for the tasks of query processing.
    There have also been studies showing that anchor text resembles real queries in terms of term distribution and length~\cite{anchor_analysis03}. Given these results, in this paper we leveraged an anchor log in a web test collection to train a deep generative model for the task of query reformulation.

    For example, given the original query ``auto insurance'', Table~\ref{tbl:auto_insurance} shows the generated queries using the proposed solution.
    Surprisingly, the model automatically learn from the users on the web to generate substitution (`\textbf{car} insurance'), addition (`\textbf{cheap} auto insurance'), and deletion (`insurance') queries.
    The most promising part of the proposed solution is that the suggested queries generated character by character. The benefits of character-level translation over word-level translation are already well known in the field of neural machine translation~\cite{full_char_nmt16}, text classification~\cite{char_conv_text_classification15} and language modeling~\cite{char_aware_nlm15}.

    \begin{table}
    \caption{Top-10 generated queries with the learned model for the input query ``auto insurance''. The generated queries includes both the original query words and new words. The score in front of each query can be considered as the importance of that query.}
    \begin{center}
    \begin{tabular}{cl}
    \toprule
    Score & Generated Queries \\ \midrule
    $0.358$ & auto insurance \\
    $0.229$ & \textbf{car} insurance \\
    $0.177$ & insurance \\
    $0.087$ & \textbf{cheap} auto insurance \\
    $0.058$ & auto insurance \textbf{quotes} \\
    $0.043$ & \textbf{car} auto insurance \\
    $0.033$ & insurance auto \\
    $0.009$ & autos insurance \\
    $0.008$ & online insurance \\
    $0.004$ & canada insurance \\
    \bottomrule
    \end{tabular}
    \end{center}
    \label{tbl:auto_insurance}
    \end{table}

    Our contribution is to adopt the character-level encoder-decoder approach of Lee et al.~\cite{full_char_nmt16} in the field of neural machine translation, and introduce completely new solution for the task of query reformulation on the web.
    The proposed solution automatically learn from the actual users on the web to reformulate the weak queries. Experimental results using standard metrics on benchmark datasets show that the generated queries achieve statistically significant improvements in retrieval effectiveness.

    The rest of the paper is organized as follows. We first review the related work in Section~\ref{sec:related_work}.
    Section~\ref{sec:approach} presents the learning model for the task of query reformulation.
    Section~\ref{sec:Training} describes the process of building the training set from the anchor log corpus and learning the model to reformulate queries. Finally, experimental results and conclusion have been presented in Section~\ref{sec:Evaluation} and~\ref{sec:conclusion} respectively.

\section{Related Work}
\label{sec:related_work}

    As mentioned in the previous section, we follow successful sequence to sequence models and adopt it to the task of query reformulation on the web. Since the proposed solution is completely novel, we separate related works in two sections. In Section~\ref{sec:query reformulation} we briefly review the related work in the task of query reformulation and Section~\ref{sec:seq2seq learning} presents a brief history of sequence to sequence learning with focus on character-level encoder-decoder models.

    \subsection{Query Reformulation}
    \label{sec:query reformulation}

    Query reformulation has a long history, dating back to the earliest explicit/pseudo relevance feedback techniques where queries are modified based on documents judged to be relevant or irrelevant~\cite{ir_in_practice09,query_reformulation_anchor10}.
    In order to apply relevance feedback an extra round of retrieval is required during the initial query reformulation phase.
    Given the constraints on response time in web search engines, relevance feedback introduces a significant extra delay to query processing~\cite{robustqr13}.

    Increasing query response time leads researchers to adjust (rewrite/expand/reformulate/suggest) the input query based on large-scale linguistic resources.
    Therefore, recent query reformulation techniques have exploited external sources such as query logs~\cite{mining_term_ass08,lambda_merge11,qr_joint_modeling15,jones06,qr_strategies09} or anchor texts~\cite{query_reformulation_anchor10,mining_anchor_qr04,robustqr13}.

    These repositories provide valuable sources of information on how users formulate the same information need on the web~\cite{ir_verbose_queries15}.
    Independent of the type of the solution, these techniques
    have focused on generating related words and phrases to substitute or expand the original query. By substituting a word in the original query with another, we run a risk of changing the users' original intent~\cite{query_reformulation_anchor10}.

    Perhaps the work that is closest to our end goal is Xue and Croft~\cite{modeling_qr13}, which transform the original query into a distribution of actual reformulated queries. This approach considers an actual query as the basic unit and thus captures important query-level dependencies between words and phrases.

    In this paper, we present an end-to-end deep learning model which is capable to learn from the actual users on the web to formulate the weak queries. Despite previous fragile term-based approaches, since the proposed solution operate on the query level, it is robust on the risk of changing the users' original intent.

    \subsection{Sequence to Sequence Learning}
    \label{sec:seq2seq learning}

    Sequence to sequence models~\cite{seq2seq14} aims at building an end-to-end deep neural network that takes as input a {\em sequence} $X=(x_{1},\ldots, x_{n})$ and returns a {\em sequence} $Y=(y_{1},\ldots, y_{m})$, where $x_t$ and $y_{t^{\prime}}$ are source and target symbols respectively, while $n$ and $m$ are the length of the source and target sequence.
    This approach has delivered state of the art performance in areas such as neural machine translation in both industry~\cite{gnmt16,neural_transliteration16} and  academia~\cite{nmt_attention15,nmt_properties14,learning_phrase_rnn14,bytenet16}.

    The important property of sequence to sequence model is that the network is oblivious to the nature of the tokens. During the training process, the semantics of the tokens are simply learned to maximize the translation quality.
    Despite the success of word-level neural machine translation~\cite{rare_words_nmt15,nmt_properties14,learning_phrase_rnn14,nmt_attention15}, recent researches revealed the importance of character-level sequence to sequence models not only in the task of neural machine translation~\cite{full_char_nmt16,char_decoder16} but also text classification~\cite{char_conv_text_classification15} and language modeling~\cite{char_aware_nlm15}.

    One advantage of character-level natural language modeling is that it can model the composition of any character sequence, thereby better modeling rare morphological variants which are prevalent on the web queries.
    Following these successful researches~\cite{full_char_nmt16,char_decoder16,char_conv_text_classification15,char_aware_nlm15,bytenet16} and the vital benefits of modeling web queries in the level of characters, we leveraged a fully character-level encoder-decoder model~\cite{full_char_nmt16} in the task of query reformulation.

    Our model learn from the actual users on the web to formulate an effective web query. Using freely available resources to train a deep model in an unsupervised manner is a hot research topic.
    For example, Kiros et al.~\cite{skip_thought15} proposed an unsupervised approach to train a word-level encoder-decoder model~\cite{learning_phrase_rnn14} using the continuity of text from a large book corpus. Experiments on a set of natural language tasks show that sentences that share semantic and syntactic properties are thus mapped to similar vector representations. Similar model proposed by Palangi et al.~\cite{deep_sent_embd16} which is a weakly supervised encoder-decoder model on user click-through data logged by a commercial web search engine. Their model evaluated on the task of web document retrieval.
    Also, Vinyals and Le~\cite{neural_conversational_model15} leveraged the same model in the task of neural conversational agent which is trained on a large open-domain movie transcript dataset. To our knowledge, this paper is the first attempt to learn from anchor texts on the web for the task of query processing.

\section{Approach}
\label{sec:approach}

    Our work is based on a few existing approaches that applied character-level convolutional neural networks to text, most notably in neural machine translation~\cite{full_char_nmt16,bytenet16}, text classification~\cite{char_conv_text_classification15} and language modeling~\cite{char_aware_nlm15}. Exactly, we follow Lee et al.~\cite{full_char_nmt16} to formulate query reformulation as a discrete sequence prediction problem. That is, we want to predict the next character in the reformulated query given its past characters, conditioned on the input query. This model consists of two component: Encoder and Decoder.

    \subsection{Encoder}
    The encoder neural network encodes the input query $X=(x_{1},\ldots, x_{n})$ into its continuous representation that summarizes its meaning, where $x_i$ is the characters of the input query. In this paper, we closely follow the encoder model proposed in~\cite{full_char_nmt16} and use a {\em character-level convolutional neural network} with max-pooling and highway layers to reduce the length of the source query representation. The shorter representation, instead of the full character query, is passed through a bidirectional network, which consists of forward and backward recurrent neural networks.
    The forward network reads the input query in the forward direction:
    \begin{equation}
        \overrightarrow{h}_{t}=\phi(e_x(x_t), \overrightarrow{h}_{t-1})
    \end{equation}
    where $e_x(x_t)$ is a continuous embedding of the $t$-th character in the input query, and $\phi$ is a recurrent activation function (gated recurrent units~\cite{learning_phrase_rnn14} in our experiments). Similarly, the reverse network reads the input query in a reverse direction:
    \begin{equation}
        \overleftarrow{h}_{t}=\phi(e_x(x_t), \overleftarrow{h}_{t+1})
    \end{equation}
    At each character location in the input query, we concatenate the hidden states from the forward and reverse recurrent network to form a set of continuous source query representations $C=\{h_{1},...,h_{n}\}$, where $h_t = [\overrightarrow{h}_{t}; \overleftarrow{h}_{t}]$.

    \subsection{Decoder}
    The decoder computes the conditional distribution over all possible queries $Y=(y_{1},\ldots, y_{m})$ given the source query $X=(x_{1},\ldots, x_{n})$, where $x_{t}$ and $y_{t^{\prime}}$ are characters in the source and target query respectively:
    \begin{equation}\label{eq:decoder_dist}
        p(Y|X)=\sum_{t^{\prime}=1}^{m} p(y_{t^{\prime}}|y_{<t^{\prime}}, X)
    \end{equation}
    For each conditional term in Equation~\ref{eq:decoder_dist}, the decoder network updates its hidden state by:
    \begin{equation}
        h_{t^{\prime}} = \phi(e_y(y_{t^{\prime}-1}), h_{t^{\prime}-1}, c_{t^{\prime}})
    \end{equation}
    where $e_y(y_{t^{\prime}-1})$ is the continuous embedding of the previous target character. $c_{t^{\prime}}$ is a context vector created by a soft-alignment mechanism~\cite{nmt_attention15}:
    \begin{equation}
        c_{t^{\prime}}=\sum_{t=1}^{n}\alpha_{t,t^{\prime}}h_{t}
    \end{equation}
    The soft-alignment weights each vector in the context set $C$ according to its relevance given what has been translated. Following previous researches~\cite{nmt_attention15,full_char_nmt16}, we used a feedforward network with a single hidden layer in order to predict the alignment probability between the $t^{\prime}$-th target character and $t$-th source character. A recurrent neural network then takes the source context vector from the attention module and predicts each target character.

    At test time, we feed a character into the decoder and get a distribution over what characters are likely to come next. Using the probability distribution over the target characters, we can select a character by sampling the distribution, and feed it right back in to get the next character. Repeating this process leads to a set of reformulated queries for a specific input query. It's the most interesting part of the model, it can generate a number of reformulated queries for the corresponding input query.

    The whole model, consisting of the encoder, decoder and soft-alignment mechanism can be trained end-to-end by minimizing the negative conditional log-likelihood using stochastic gradient descent.

\section{Training}
\label{sec:Training}

    In this section, we train the proposed solution using readily available anchor texts on the web. Section~\ref{sec:Training Set Construction} and~\ref{sec:Training Details} represent the training set construction and the training details respectively.

    \subsection{Training Set Construction}
    \label{sec:Training Set Construction}

    \begin{figure}
    \centering
    \begin{tikzpicture}
       \GraphInit[vstyle=Normal]
       \tikzset{VertexStyle/.style   = {shape = circle, draw}}
       \Vertex[x=2,y=2, L=$u_j$]{A}
       \tikzset{VertexStyle/.style   = {shape = rectangle, draw}}
       \Vertex[x=4,y=4, L=$a_{2}$]{B}
       \Vertex[x=0,y=4, L=$a_{1}$]{C}
       \Vertex[x=0,y=0, L=$a_{3}$]{D}
       \Vertex[x=4,y=0, L=$a_{4}$]{E}
       \tikzset{EdgeStyle/.style={post}}
       \Edge[label=$f_{2,j}$](B)(A)
       \Edge[label=$f_{1,j}$](C)(A)
       \Edge[label=$f_{3,j}$](D)(A)
       \Edge[label=$f_{4,j}$](E)(A)
    \end{tikzpicture}
    \caption{A session has been defined in the Clueweb09 anchor text query log corpus~\cite{query_reformulation_anchor10} to be simply the group of anchor texts $\{a_{1}, a_{2}, a_{3}, a_{4}\}$ that point to the same web page ($u_j$).}
    \label{fig:clueweb09_anchor_log}
    \end{figure}
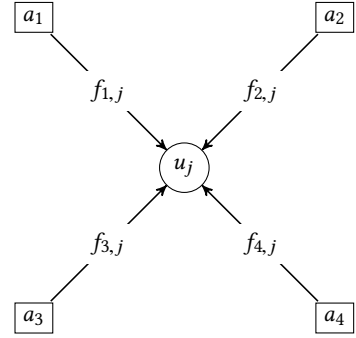

    We need a set of query pairs $\langle q, q^{r} \rangle$ in order to train the deep model presented in Section~\ref{sec:approach}, which can be leveraged by various query log mining methods in commercial web search engines. In absence of query logs, anchor text is a good approximation. In this section, using the availability of anchor texts, we present a way to automatically construct a huge training set that represent the ability of web users in expressing the same information need.

    Dang and Croft~\cite{query_reformulation_anchor10} constructed anchor text query log corpus from Clueweb09~\footnote{http://lemurproject.org/clueweb09/anchortext-querylog/} that consists of pairs $\langle u_{j}, a_{i}, f_{i,j} \rangle$ where $u_{j}$ is the associated web page for the anchor text $a_{i}$ and $f_{i,j}$ is the frequency of $\langle u_j, a_i \rangle$ in the whole corpus.
    We follow the approach of Dang and Croft~\cite{query_reformulation_anchor10}, using a large anchor graph as a linguistic resource in order to train an end-to-end deep model for the task of query reformulation on the web.

    As shown in Figure~\ref{fig:clueweb09_anchor_log}, we defined a session in the anchor log to be simply the group of anchor texts $\{a_{1}, a_{2}, \ldots, a_{n}\}$ that point to the same web page $u_j$.
    This set of anchors provides valuable sources of information on how users formulate/reformulate the same information need in different syntactic or semantic ways.

    Suppose that among the set of anchor texts $\{a_{1}, a_{2}, \ldots, a_{n}\}$ associated to a specific web page, $a^{*}$ in the most frequent anchor among users on the web. In a real world scenario, $a_{i}$ could be different ways of expressing $u_j$ contains misspelling and rare words, while $a^{*}$ is the most common anchor in expressing $u_j$ among the users on the web. On the other hand, it is possible to map each ill-formed anchor to the corresponding well-known anchor on the web ($a_{i}\xmapsto[]{}a^{*}$).

    A large portion of anchor texts are automatically generated by web authoring tools, or are for other reasons useless.
    We need to filter the irrelevant anchor texts to obtain a list of high quality anchor texts that are similar and semantically related to the original query. We follow simple rules to filter $\langle u_{j}, a_{i}, f_{i,j} \rangle$ recorrds in the anchor log corpus:
    \begin{itemize}
        \item $f_{i,j} < 2$: The occurrences of anchor $a_{i}$ to the page $u_{j}$ is less than two times in the whole corpus.
        \item $| a_{i} | < 50$: The number of characters in the anchor text is less than $50$, since the average length of the queries on the web is $2.21$ words~\cite{web_search_analysis00}.
        \item $\frac{ W_{a_{i}} \cap W_{a^{*}} } { W_{a_{i}} \cup W_{a^{*}} } < 0.3$: Where $W_{a_{i}}$ is the set of words in the anchor text $a_{i}$. In this way the well-known anchor $a^{*}$ is strongly related to the original anchor $a_{i}$, containing common terms.
    \end{itemize}

    Anchor text is often a concise representation of the page it points to and can be thought of as a `query' to retrieve that page~\cite{query_reformulation_anchor10}. Since the anchor text is chosen manually to represent the page, its content is very relevant to the corresponding document and often provide more accurate description of the page than the page itself. The similarity between anchor text and real queries has also been observed by other researchers~\cite{query_reformulation_anchor10,anchor_analysis03}.
    If we relate each anchor text to a query, it is possible to construct a huge set of query reformulation pairs. Following the above rules, we construct a large set of query pairs $\langle q, q^{r} \rangle$ containing 138,107,683 records. A small set containing 1000 random records leveraged as the validation set in the following experiments.

    By making use of anchor log for training an end-to-end deep model, we obtain a number of significant benefits~\cite{mining_anchor_qr04}. First, this data is freely available and there is no need for laborious labeling which is the most important barrier in supervised learning. On the other hand, by mining anchor log, we trained our supervised model in an unsupervised manner. Second, the use of anchor texts as training data for the task query reformulation provide valuable sources of information on how users formulate/reformulate the same information needs on the web in different syntactic or semantic ways. So the model have the possibility to learn these syntactic/semantic patterns of reformulations.

    \subsection{Training Details}
    \label{sec:Training Details}

    \begin{figure}
        \centering
        \includegraphics[width=0.364\textwidth]{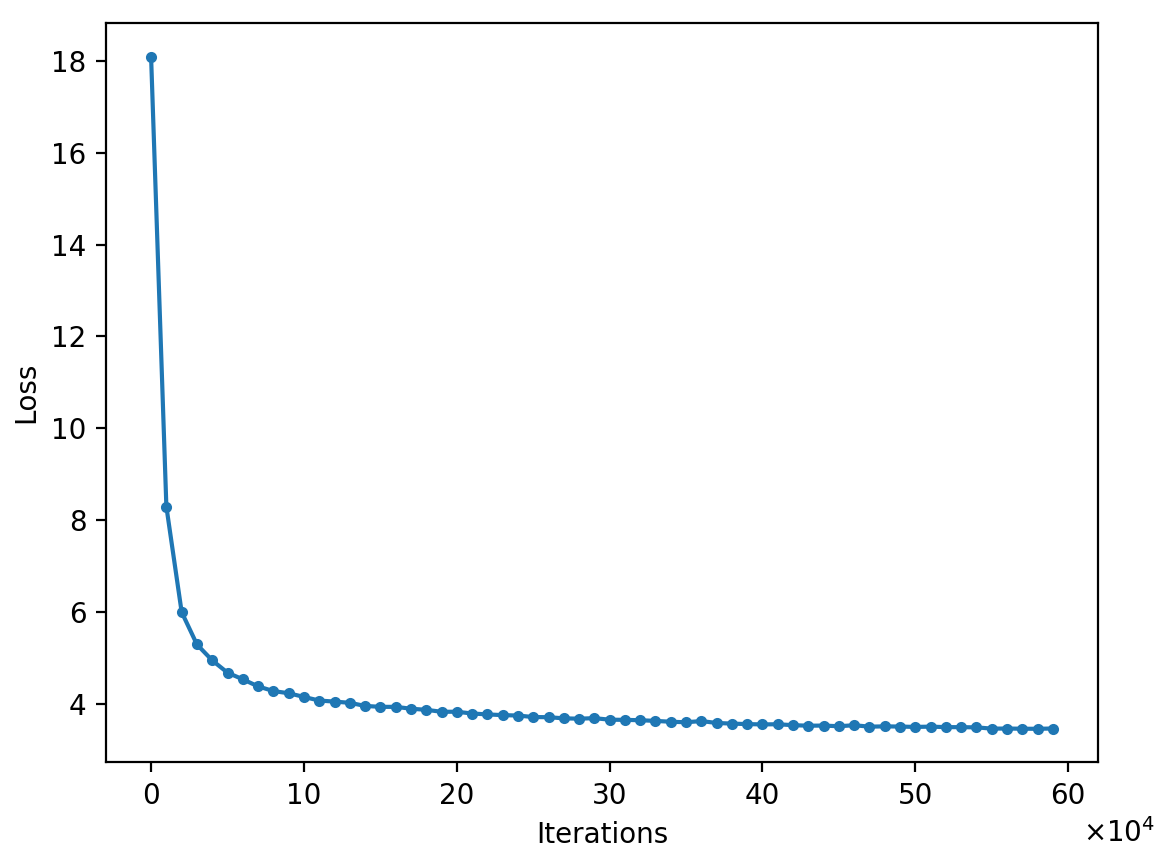}
        \caption{Validation loss vs. iterations during the learning process. The validation set contains $1000$ random samples which are removed from the training set.}
        \label{fig:learning_curve}
    \end{figure}

    Following previous research~\cite{full_char_nmt16}, the model is trained using stochastic gradient descent using the Adam optimization~\cite{adam14}, learning rate $0.0001$ and minibatch size $64$. The norm of the gradient is clipped with a threshold of $1.0$. All weights are initialized from a uniform distribution $[-0.01, 0.01]$.

    Figure~\ref{fig:learning_curve} shows the validation curve during the learning process. The horizontal and vertical axes represent the learning iterations and the validation loss respectively. As mentioned in Section~\ref{sec:Training Set Construction}, the validation set contains $1000$ random samples which are removed from the training set.

    For decoding, a two-layer character-level attentional decoder with $1024$ Gated Recurrent Units~\cite{nmt_properties14} is used in the following experiments. We use beam search as a sampling method with length-normalization while beam width set to $30$.

\section{Evaluation}
\label{sec:Evaluation}

    After training the model, we can evaluate the effectiveness of the learned model in the task of query reformulation using benchmark data sets and standard metrics.
    We compare the retrieval performance given by the proposed reformulation method trained on anchor text query log corpus against the baseline original queries.

    \subsection{Experimental Setup}
    We used Lemur/Indri~\cite{ir_in_practice09} as the retrieval software.
    Indri's language modeling retrieval engine~\cite{indri04,indri05} with default parameters was used for the baseline retrieval.
    For each query, a ranking of the top $1000$ documents has been retrieved~\footnote{ Participants in the adhoc task of TREC 2010-2012~\cite{trec10web,trec11web,trec12web} submitted a ranking of the top $10000$ documents for each topic.}.
    We set $m$ (the number of generated candidates considered by the reformulation model) to $10$ in the following experiments.

    \subsubsection{Test Collection}
    In order to simulate the context of web search, all experiments is done using Clueweb09 Category B dataset~\footnote{http://lemurproject.org/clueweb09/}.
    It consists of 50 million English web pages with large and varied vocabulary.
    The corpus was processed with the default stoplist, and stemmed using the Krovetz stemmer.

    \subsubsection{Query Set}
    There are four query sets (TREC 2009-2012) associated to the Clueweb09 dataset.
    The TREC 2009 web Track~\cite{trec09web} included a traditional adhoc retrieval task while TREC 2010-2012~\cite{trec10web,trec11web,trec12web} incorporated multiple relevance levels, which are similar in structure to the levels used in a web search engine.
    Each query set contains a set of $50$ topics.
    Each topic contains a query field, a description field, and several subtopic fields.
    The query field is intended to represent the text a user might enter into a web search engine.
    In order to have a uniform evaluation, we used all $150$ query fields associated with TREC 2010-2012~\cite{trec10web,trec11web,trec12web} in the following experiments.

    \subsubsection{Evaluation Metric}
    The primary effectiveness measure for the adhoc task in TREC 2010-2012~\cite{trec10web,trec11web,trec12web} is expected reciprocal rank (ERR)~\cite{err}.
    We also report a variant of normalized discounted cumulative gain (nDCG), as well as standard binary measures, including mean average precision (MAP) and precision at rank $k$ (P@$k$).

    \subsection{Experiment Design}
    \label{sec:experiment_design}

    Previous researches~\cite{query_reformulation_anchor10} revealed that it is difficult to evaluate query reformulation techniques since the definition of ``quality'' of reformulations is unclear. According to the previous studies~\cite{query_reformulation_anchor10,robustqr13}, we choose to evaluate reformulations using their retrieval performance.

    At query time, for each query $q$, the top $m$ candidates given by the learned reformulation model $\{ q_{1}^{r},\ldots, q_{m}^{r} \}$ were used.
    The best ERR@20 obtained by these $m$ candidates is recorded as the ERR@20 of the reformulation solution for $q$.
    We varied $m$ from $1$ to $10$ in our experiment.
    This can be explained in the context of an end user using a web search engine when the system suggests a list of reformulations to the user~\cite{when_query_suggest13}.

    \subsection{Results}
    \label{sec:results}

    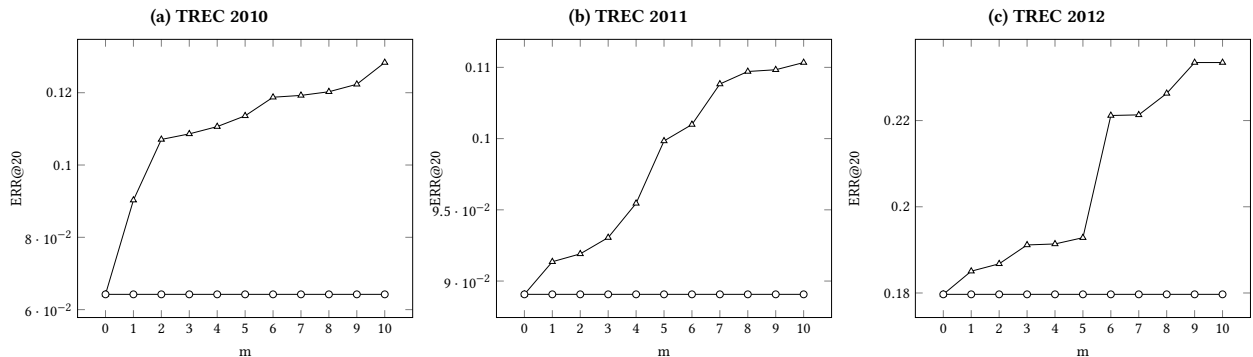
\begin{figure*}
    \centering
    \subfloat[TREC 2010]{
        \begin{tikzpicture}[scale=0.65]
            \begin{axis}[
                xlabel=m,
                ylabel=ERR@20,
                xtick={0, 1, 2, 3, 4, 5, 6, 7, 8, 9, 10}]
            \addplot[mark=triangle*, mark options={fill=white}] plot coordinates {
                (0, 0.0642246)
                (1, 0.0902846)
                (2, 0.10707260000000002)
                (3, 0.10861140000000002)
                (4, 0.11062460000000002)
                (5, 0.11360480000000002)
                (6, 0.11874920000000001)
                (7, 0.11925540000000001)
                (8, 0.12026580000000003)
                (9, 0.12233440000000001)
                (10, 0.12834580000000004)
            };
            \addplot[mark=*, mark options={fill=white}] plot coordinates {
                (0, 0.0642246)
                (1, 0.0642246)
                (2, 0.0642246)
                (3, 0.0642246)
                (4, 0.0642246)
                (5, 0.0642246)
                (6, 0.0642246)
                (7, 0.0642246)
                (8, 0.0642246)
                (9, 0.0642246)
                (10, 0.0642246)
            };
        \end{axis}
        \end{tikzpicture}
    }
    \subfloat[TREC 2011]{
        \begin{tikzpicture}[scale=0.65]
            \begin{axis}[
                xlabel=m,
                ylabel=ERR@20,
                xtick={0, 1, 2, 3, 4, 5, 6, 7, 8, 9, 10}]
            \addplot[mark=triangle*, mark options={fill=white}] plot coordinates {
                (0, 0.08906739999999998)
                (1, 0.09135599999999998)
                (2, 0.09190839999999997)
                (3, 0.0930526)
                (4, 0.0954546)
                (5, 0.0998348)
                (6, 0.10098860000000003)
                (7, 0.10383040000000004)
                (8, 0.10471000000000004)
                (9, 0.10483320000000004)
                (10, 0.10533900000000004)
            };
            \addplot[mark=*, mark options={fill=white}] plot coordinates {
                (0, 0.08906739999999998)
                (1, 0.08906739999999998)
                (2, 0.08906739999999998)
                (3, 0.08906739999999998)
                (4, 0.08906739999999998)
                (5, 0.08906739999999998)
                (6, 0.08906739999999998)
                (7, 0.08906739999999998)
                (8, 0.08906739999999998)
                (9, 0.08906739999999998)
                (10, 0.08906739999999998)
            };
        \end{axis}
        \end{tikzpicture}
    }
    \subfloat[TREC 2012]{
        \begin{tikzpicture}[scale=0.65]
            \begin{axis}[
                xlabel=m,
                ylabel=ERR@20,
                xtick={0, 1, 2, 3, 4, 5, 6, 7, 8, 9, 10}]
            \addplot[mark=triangle*, mark options={fill=white}] plot coordinates {
                (0,0.17968600000000007)
                (1,0.18505660000000002)
                (2,0.1867676)
                (3,0.191143)
                (4,0.19137500000000002)
                (5,0.1928294)
                (6,0.2211734)
                (7,0.22131919999999997)
                (8,0.22629619999999995)
                (9,0.23345839999999995)
                (10,0.23345839999999995)
            };
            \label{mark:qr}
            \addplot[mark=*, mark options={fill=white}] plot coordinates {
                (0, 0.17968600000000007)
                (1, 0.17968600000000007)
                (2, 0.17968600000000007)
                (3, 0.17968600000000007)
                (4, 0.17968600000000007)
                (5, 0.17968600000000007)
                (6, 0.17968600000000007)
                (7, 0.17968600000000007)
                (8, 0.17968600000000007)
                (9, 0.17968600000000007)
                (10, 0.17968600000000007)
            };
            \label{mark:lm_baseline}
        \end{axis}
        \end{tikzpicture}
    }
    \caption{Performance comparison among the original queries in TREC 2010-2012 query sets~\cite{trec10web,trec11web,trec12web} and their top-$m$ reformulations generated with the trained model. The parameter $m$ represents the number of generated candidates considered by the reformulation model and ~\ref*{mark:lm_baseline},~\ref*{mark:qr} indicate the retrieval performance of the baseline Indri language model using the original query and the proposed reformulated query respectively.}
    \label{fig:qr_results}
    \end{figure*}

    Figure~\ref{fig:qr_results} compares the performance of the generated queries and the original input query.
    Given $m$ reformulated queries for each original query, we select the best one which has the best ERR@20.
    This can be explained in the web search context when an end user capable of selecting the most informative query among a suggested list of reformulated queries automatically generated by the proposed solution~\cite{when_query_suggest13}.

    Form Figure~\ref{fig:qr_results} we can see that the learned reformulation model  significantly outperforms original queries on all TREC collections~\cite{trec10web,trec11web,trec12web} and thus can recommend more meaningful queries.
    When we consider up to top 10 reformulations for each original query, the proposed solution provides a relative improvement of 100\%, 18\% and 30\% on TREC 2010~\cite{trec10web}, 2011~\cite{trec11web} and 2012~\cite{trec12web} respectively, if the best query formulation is used for each query. It's clear that the trained model learned syntactic/semantic patterns of reformulations using more than 100 millions of reformulation patterns on the web, is capable of generating more informative queries than the original input query.

    \begin{table*}
    \begin{center}
    \begin{tabular}{lcccccccccc}
    \toprule
    & & \multicolumn{4}{c}{Original Queries} & &
    \multicolumn{4}{c}{Best Reformulated Queries} \\
    \cmidrule{3-6}  \cmidrule{8-11}
    & & MAP & P@20 & nDCG@20 & ERR@20 & &
    MAP & P@20 & nDCG@20 & ERR@20 \\
    \midrule
    TREC 2010 & & $0.0930$ & $0.2000$ & $0.1091$ & $0.0669$ & & $0.0984$ & $\mathbf{0.2667}$ & $\mathbf{0.1745}$ & $\mathbf{0.1337}$ \\
    TREC 2011 & & $0.0921$ & $0.2390$ & $0.1583$ & $0.0891$ & & $0.0906$ & $0.2360$ & $\mathbf{0.1724}$ & $\mathbf{0.1053}$ \\
    TREC 2012 & & $0.0912$ & $0.1970$ & $0.0971$ & $0.1797$ & & $0.0916$ & $0.2040$ & $\mathbf{0.1200}$ & $\mathbf{0.2335}$ \\
    \bottomrule
    \end{tabular}
    \end{center}
    \caption{Performance comparison among the original queries in TREC 2010-2012 collections~\cite{trec10web,trec11web,trec12web} and the best reformulated query proposed by the learned model.}
    \label{tbl:qr}
    \end{table*}

    The most important aspect of Lee et al.~\cite{full_char_nmt16} sequence to sequence model is the absence of explicitly providing knowledge of words. On the other hand, the model learns it character by character. Despite the reformulated queries generate character by character in our experiments, we found that the learned model completely aware of the word's concepts and their boundaries. The alphabet used in our experiments consists of 41 characters, including 26 english letters, 10 digits, 5 other special characters.

    Table~\ref{tbl:qr} shows the performance of the original input query and the best reformulated query proposed by the learned reformulation model using nDCG at rank $20$ (nDCG@$20$), as well as standard binary measures, including mean average precision (MAP) and precision at rank $20$ (P@$20$). The experiment has been done on all available query sets in TREC 2010-2012~\cite{trec10web,trec11web,trec12web}.



    \begin{table*}
    \caption{Examples of good generated queries with the learned model. The first column shows the original query of which ERR@20 is given in the second column. The suggestions to this query together with its ERR@20 is provided in the third and fourth column respectively.}
    \begin{center}
    \begin{tabular}{llclc}
    \toprule
    QID & Original Query & ERR@20 & Reformulated Query & ERR@20 \\ \midrule
    90  & mgb & $0.0289$ & mgb \textbf{motor} & $0.3749$ \\
    77  & bobcat & $0.0625$ & bobcat \textbf{company} & $0.5368$ \\
    159 & porterville & $0.1639$ & porterville \textbf{city} & $0.9681$ \\
    135 & source of the nile & $0.0000$ & \textbf{the} source of the nile & $0.0094$ \\
    137 & \textbf{rock and} gem shows & $0.1755$ & gem shows & $0.2093$ \\
    139 & rocky \textbf{mountain} news & $0.0057$ & rocky news & $0.0182$ \\
    185 & credit report & $0.1875$ & \textbf{free} credit report & $0.3125$ \\
    59  & how to build \textbf{a} fence & $0.1144$ & how to build fence & $0.4447$ \\
    \bottomrule
    \end{tabular}
    \end{center}
    \label{tbl:examples}
    \end{table*}

    Table~\ref{tbl:examples} gives some examples of good query reformulations provided by our solution.
    All these reformulated queries dramatically improve the performance of the corresponding original query.
    For some queries, the learned model adds context words to better represent the intent of the user (such as: `mgb +[motor]', `bobcat +[company]' or `porterville +[city]').
    For some other queries, the learned reformulation model generates a well-known query for the corresponding input query (such as: `gem shows' or `rocky news').
    Surprisingly, the learned model add/remove some function words to/from the original query which significantly improve the retrieval performance (such as: `[the]+ source of the nile' or `how to build [a]- fence'). This result, however, is not unexpected since our decoder is a neural language model which conditions on a set of encoder output $h_{t}$ and the Indri retrieval model~\cite{indri05} combines the language modeling and inference network approaches to information retrieval.

    \subsection{Failure Analysis}
    \label{sec:Failure Analysis}

    \begin{table}
    \caption{Top-10 generated queries for the input query ``grilling''. Ambiguous query number 160 from TREC 2012~\cite{trec12web}: ``Find kabob recipes, tips on grilling vegetables and fish, instructions for grilling chicken''.}
    \begin{center}
    \begin{tabular}{cl}
    \toprule
    Score & Generated Queries \\ \midrule
    $0.625$ & grilling \\
    $0.160$ & grilling \textbf{real estate} \\
    $0.099$ & www grilling com \\
    $0.094$ & grilling \textbf{hotels} \\
    $0.060$ & grilling \textbf{real estage} \\
    $0.049$ & about grilling \\
    $0.041$ & the grilling \\
    $0.023$ & grilling s \textbf{blog} \\
    $0.021$ & grilling schools \\
    $0.005$ & grilling records \\
    \bottomrule
    \end{tabular}
    \end{center}
    \label{tbl:grilling}
    \end{table}

    Although the learned model generally performs well, after analysing the generated queries for all 150 topics in TREC 2010-2012~\cite{trec10web,trec11web,trec12web}, some failures are also observed.
    We found that the generated queries in some cases contains popular phrases such as: `real estate', `hotels', `blog' and `services' while it does not relate to the original query. Table~\ref{tbl:grilling} shows top-10 generated queries for the input query `grilling'. The term `real estate' and `hotels' occurs in about 322K and 889K training samples respectively while `grilling' only occurs in about 1K of training samples.

    Previous researches~\cite{clue09_spam11} revealed the importance of spam filtering for Clueweb09 corpus. As mentioned in Section~\ref{sec:Training Set Construction}, we leveraged all available anchor texts from the 500 million English web pages in the Clueweb09 anchor log~\cite{query_reformulation_anchor10}.
    Consider the popularity of these phrases between web spammers, it seems that anchors coming from link spams leads to these negative results.

\section{Conclusions and future work}
\label{sec:conclusion}

In this paper, we present a solution for the task of web query reformulation following state-of-the-art approaches in neural machine translation~\cite{full_char_nmt16}. Compare to the classic methods in the field on query reformulation, the most important strength of our solution is that it can be trained end-to-end and thus requires much fewer hand-crafted rules to cover different types of query reformulation strategies~\cite{qr_strategies09}. By training an end-to-end query to query model instead of substituting the original query word with another, the proposed solution robust on the risk of changing the users' original intent.

From the view point of deep learning models, the most important aspect of the proposed solution is that it leveraged freely available anchor texts to learn from actual users on the web to reformulate queries. On the other hand, we proposed a way to train a deep model for the task of query reformulation in an unsupervised manner.
Despite the successful usage of anchor texts in our experiments, the proposed solution can be trained on other valuable resources such as query logs.

Our experiments show that the reformulated queries generated by the proposed solution significantly improve the retrieval performance using standard measures on benchmark TREC datasets. We leveraged a state-of-the-art character-level sequence to sequence model~\cite{full_char_nmt16} in order to not only understand the original input query but also generate effective reformulated queries character by character. Leveraging characters instead of words have a lot of benefits in modeling web queries.
Dang and Croft~\cite{query_reformulation_anchor10} show that reformulation techniques that worked well on ill-formed web queries do not significantly improve well-formed TREC queries. Therefore, the potential benefit of the proposed approach in modeling ill-formed web queries remains intact for future works.

\begin{acks}
    The authors would like to thank the developers of Theano deep learning framework~\cite{theano16} and the developers of dl4mt-c2c project~\footnote{https://github.com/nyu-dl/dl4mt-c2c}. This research was not possible without their contribution.
    The author also appreciate Prof. Jamie Callan and the developers of the Lemur project for getting access to Batch Query Service for Clueweb09~\footnote{http://boston.lti.cs.cmu.edu/Services/clueweb09\_batch/}.
\end{acks}

\bibliographystyle{ACM-Reference-Format}
\bibliography{references}

\end{document}